\documentclass[preprint,preprintnumbers,amsmath,amssymb]{revtex4}

\usepackage{graphicx}
\usepackage{bm}

\begin{document}

\preprint{CAS-KITPC/ITP-090}
\title{ ~\\ \vspace{2cm}
Multi-Stream Inflation\vspace{1cm}}

\author{Miao Li}\email{mli@itp.ac.cn}
\author{Yi Wang}\email{wangyi@itp.ac.cn}
\affiliation{Kavli Institute for Theoretical Physics China,
Key Laboratory of Frontiers in Theoretical Physics,
Institute of Theoretical Physics, Chinese Academy of Sciences,
Beijing 100190, P.R.China\vspace{2cm}}

\begin{abstract}
We propose a ``multi-stream'' inflation model, which is a double
field model with spontaneous breaking and restoration of an
approximate symmetry. We calculate the density perturbation and
non-Gaussianity in this model. We find that this model can have
large, scale dependent, and probably oscillating non-Gaussianity. We
also note that our model can produce features in the CMB power
spectrum and hemispherical power asymmetry.
\end{abstract}

\maketitle

\section{Introduction}

Inflation has become one of the most important windows to connect
cosmological observations to fundamental physics, such as string
theory. Recent studies on string theory show evidence that inflation
may happen in the string theory landscape \cite{Bousso:2000xa} with
extremely complicated inflaton potential \cite{Huang:2008jr}.

One lesson from the string landscape is that the inflationary
dynamics may be not as simple as one expected before, and more
possibilities during inflation should be taken into consideration.
Inflation in the string landscape opens up a number of possibilities
to test string theory by the stringy imprints on the CMB temperature
fluctuations.

On the other hand, in the recent analysis of CMB anisotropy, there
are several hints for new physics. If future experiments confirm
some of these possibilities, they will pose challenges for
inflationary cosmology and opportunities for string theory. These
challenges and opportunities include:

\begin{itemize}
  \item Non-Gaussianity. The WMAP5 bound for local non-Gaussianity is
  \cite{Komatsu:2008hk}
  \begin{equation}
    -9<f_{NL}<111~ (95\% {\rm CL})~.
  \end{equation}
  The result is still consistent with $f_{NL} \simeq 0$.
  However, the central value of $f_{NL}$ is large and
  positive. If integration of more years of WMAP or the Planck mission
  confirm this result, it would rule out the simplest inflation
  model. The theoretical investigation of non-Gaussianity include
  \cite{Wise,Hodges,Salopek,Srednicki,Gangui,KomatsuSpergel,Bartolo,multifield,Tong,Maldacena,Chen,Komatsu,NGrecent,aalok}.
  As there is a consistency relation \cite{Maldacena, Li:2008gg} for single field
  inflation, a large and local non-Gaussianity shows evidence that inflation
  should involve more than one field.

  \item Features in the CMB. As shown in \cite{Covi:2006ci}, there
  may be features in the CMB scalar power spectrum. A feature in
  the CMB spectrum implies a departure from the standard single field
  slow roll inflation at a certain scale.

  \item Hemispherical power asymmetry. As shown in \cite{Erickcek:2008sm},
   the temperature-fluctuation amplitude is larger
by roughly 10\% in one hemisphere of the CMB map than in the other.
Again, this asymmetry can not be explained by the simplest inflation
model. It would also be interesting to see whether this difference
of power also happens on smaller scales in the sky.
\end{itemize}

In this paper, we propose a ``multi-stream'' inflation model, which
is a double field inflation model with spontaneous breaking and
restoration of an approximate symmetry. We find that with
appropriate choice of parameters, our model is consistent with
current observations, and can produce large non-Gaussianity,
features in the CMB and hemispherical power asymmetry. In Section 2,
we describe the picture of multi-stream inflation and calculate the
density perturbations. In Section 3, we calculate the
non-Gaussianity of multi-stream inflation. In Section 4, we give a
summary of the signatures for multi-stream inflation in different
parameter regions. We conclude in Section 5.

\section{Multi-Stream Inflation}
In this section, we investigate the background evolution and the
scalar perturbation spectrum of multi-stream inflation. We consider
a period of inflation involving two fields.

\begin{figure}
  \center
  \includegraphics[width=0.6\textwidth]{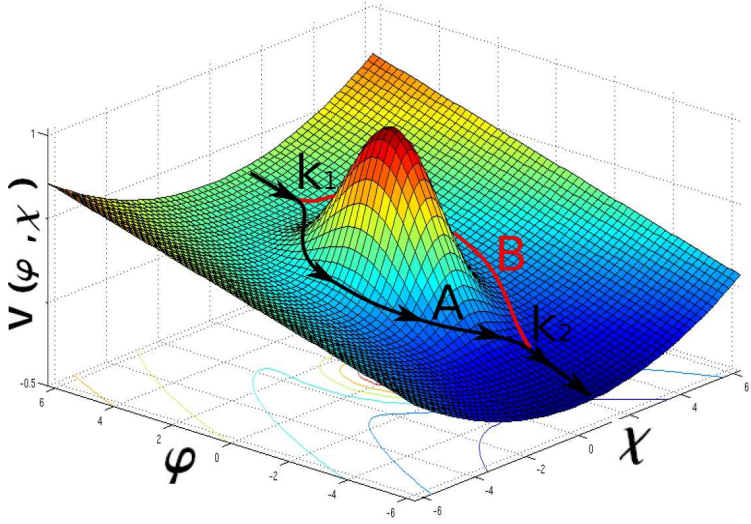}
    \caption{\label{fig:multi-stream} In this figure, we illustrate the picture of multi-stream inflation.
    As the inflaton rolls down the potential, it experiences a spontaneously
    breaking of approximate symmetry at $k_1$, with $\varphi=\varphi_1$. After that, the inflation
    trajectory branches off into two paths denoted by $A$ and $B$.
    From the viewpoint of the CMB map, some parts of the sky with size of order $1/k_1$
    come from path $A$, and other parts of the sky come from $B$.
    At $k_2$, with $\varphi=\varphi_2$, the approximate symmetry is restored and the inflationary
    direction becomes unique again.}
\end{figure}

We consider the potential $V(\varphi,\chi)$ as illustrated in
Fig.\ref{fig:multi-stream}. In the beginning, the $\varphi$
direction is the inflationary direction. As $\varphi$ evolves to
$\varphi_1$, the $\chi$ direction becomes a tachyonic direction. At
this point, the inflationary trajectory spontaneously breaks into
two paths, namely $A$ and $B$ in Fig.\ref{fig:multi-stream}. The
whole picture looks like a stream which splits and flows along both
sides of a hill. So we call this scenario ``multi-stream''
inflation.

As inflation continues, depending on the potential and the reheating
time, the trajectories $A$ and $B$ may either recombine into a
single trajectory (as shown in Fig.\ref{fig:multi-stream}) or not.
As we will see in the calculation of scalar perturbations, if $A$
and $B$ do not recombine and reheat separately, one typically has
large anisotropy at the symmetry breaking scale, and requires more
fine tuning to fit the CMB power spectrum. So in this note, we
mainly discuss the former possibility.

Now let us investigate the perturbations. The perturbations of
multi-stream inflation should be the same as in usual inflation
models except in the regime between $\varphi_1$ and $\varphi_2$.
Moreover, if the potential energy of path $A$ and path $B$ are
identical, multi-stream inflation also does not differ much from
usual inflation models except for some amount of isocurvature
perturbations, because in this case, to choose which way is not
important.

To have more interesting physics, we assume the paths $A$ and $B$
have a slightly different potential energy $\delta V\equiv
V(\varphi,\chi_A) -V(\varphi,\chi_B)$, with $\delta V$ non-vanishing
only if $\varphi_1<\varphi <\varphi_2$. After this setup,
multi-stream inflation will have several interesting differences
from the usual inflation model.

Firstly, the perturbations along trajectories $A$ and $B$ have
different power spectra. It is because the tilt of the potential is
typically different when $\delta V \neq 0$. To be more explicit, the
power spectra along the two trajectories are calculated using the
local Hubble parameter and the local slow roll parameters as
\begin{equation}
  P_\zeta^A=\left.\frac{H^2}{8\pi^2M_p^2\epsilon}\right|_A \qquad P_\zeta^B=
  \left.\frac{H^2}{8\pi^2M_p^2\epsilon}\right|_B~.
\end{equation}
The difference in the power spectra satisfies
\begin{equation}\label{PV}
(P_\zeta^A-P_\zeta^B)/P_\zeta^A \sim \delta V/V~.
\end{equation}

In position space, trajectories $A$ and $B$ correspond to different
parts of the sky. So a difference between the power spectra
$P_\zeta^A$ and $P_\zeta^B$ implies that on cosmological scales
corresponding to $\varphi_1$ and $\varphi_2$, the perturbations in
some regions of the sky have more power, and in some other regions
of the sky have less power.

For example, if $k_1$ corresponds to the scale of the whole observable
universe, then trajectories $A$ and $B$ both span about half of the
sky. In this case, multi-stream inflation provides a possible solution
to the hemispherical power asymmetry problem \cite{Erickcek:2008sm}.

If the hemispherical power asymmetry is really there and with a
difference of order $10\%$ in magnitude, we need $\delta V/V\sim
10^{-2}$, and $k_1$ to be the scale that comparable to the present
Hubble scale to explain it. As shown in the following sections, in
this case, the corresponding non-Gaussianity is typically
$f_{NL}\sim {\cal O}(100)$.

In multi-stream inflation, $k_1$ could also correspond to much smaller
length scales compared with the present Hubble scale, depending on the
explicit
form of the inflationary potential. In this case, the multi-stream
effect generates about $(k_1/k_0)^2$ regions on the CMB, where $k_0$
is the comoving wave number whose wavelength is the whole observable
universe. In these regions, the primordial power spectrum either takes
the form $P^A_\zeta$ or $P^B_\zeta$.

Secondly, there can be a difference of e-folding numbers for paths
$A$ and $B$. This difference results in a time delay effect for the
reheating surface, thus producing extra perturbation.

To be explicit, let $\delta N\equiv N_A- N_B$, where $N_A$ and $N_B$ are the
background e-folding numbers evaluated along trajectories $A$ and $B$ respectively. We note
that $\delta N$ depends on the detailed form of the potential
\begin{equation}
  \delta N=\frac{1}{M_p^2}\int_{\varphi_1}^{\varphi_2}d\varphi \frac{V+\delta V}{
  V'+\delta V'}-\frac{1}{M_p^2}\int_{\varphi_1}^{\varphi_2}d\varphi \frac{V}{
  V'}~,
\end{equation}
where a prime denotes a derivative with respect to $\varphi$. We
denote the comoving scale that exits the Hubble radius when
$\varphi=\varphi_1$ by $k_1$, and $\varphi=\varphi_2$ by $k_2$ as
shown in Fig.\ref{fig:multi-stream}. $\delta N$ provides an extra
source of curvature perturbation on the scale $k_1$. It is because
the existence of $\delta N$, thus the time delay in reheating
divides the CMB sky into patches with comoving length scale $1/k_1$.
This anisotropy corresponds to a perturbation $\delta
\zeta_{k_1}^N$. The amount of perturbation can be calculated as \cite{Sasaki:1995aw}
\begin{equation}
\zeta_{k_1}=\Delta N=\frac{H}{\dot\varphi}\delta\varphi+\delta N
\equiv \zeta_0+\delta\zeta_{k_1}^N~,
\end{equation}
where $\Delta N$ is the total e-folding number difference between
path $A$ and path $B$. The difference $\Delta N$ has two sources,
namely the quantum fluctuation of the inflation $\delta\varphi$, and
the intrinsic difference in the e-folding number $\delta N$ between
the two paths.


The total power spectrum takes the form
\begin{equation}\label{Prough}
  P_{\zeta_{k_1}}\sim \zeta_0^2+\left(\delta\zeta^N_{k_1}\right)^2~.
\end{equation}
We neglected the cross term $\zeta_0\delta\zeta_{k_1}^N$, because if
one does not consider the conversion from isocurvature perturbation
to curvature perturbation, then $\zeta_0$ and $\delta\zeta_{k_1}^N$
are uncorrelated. We will return to this issue and give a more
careful analysis of Eq. \eqref{Prough} in the next section.

The existence of $\delta N$ is a mixed blessing. The good news is
that $\delta N$ provides a nice explanation for the features in the
CMB power spectrum, because the jump of $\zeta$ takes place only in
a very sharp range of $k$. The bad news is that when $\delta V$ is
not very small, to make this jump not too large, one needs some
amount of tuning to be consistent with experiments. We shall
estimate the amount of tuning in Section 4.

\section{Non-Gaussianity of Multi-Stream Inflation}

In this section, we calculate the non-Gaussianity of multi-stream
inflation. We shall neglect the non-Gaussianity produced by
mechanisms other than the multi-stream effect.

To perform the calculation, we further assume that the mass of
$\chi$ shortly before the Hubble exit of $k_1$ is of order $H$. This
assumption guarantees that $m_\chi$ is not too light to randomly
walk more than one e-fold away from the origin, and also guarantees
that $m_\chi$ is not too heavy to have an effect.

As the inflation trajectory turns at $\varphi_1$, a certain amount
of isocurvature perturbation $\delta\chi_{k_1}$ is projected to
curvature perturbation. Note that the sign of $\delta\chi_{k_1}$
decides whether inflation will be along trajectory $A$ or $B$. The
observed curvature perturbation at $k_1$ also includes a
contribution from $\delta N$, we have
\begin{equation}
  \zeta_{k_1}\equiv \zeta_0+\delta\zeta_{k_1}~,\qquad \delta\zeta_{k_1}\equiv
  c \frac{H}{\dot\varphi}\delta\chi_{k_1}+\delta N
  \equiv\delta\zeta_{k_1}^S
+\delta\zeta_{k_1}^N~,
\end{equation}
where $c$ denotes the fraction of isocurvation which is projected
onto the curvature perturbation during the change of the trajectory,
$\delta\zeta_{k_1}^S$ and $\delta\zeta_{k_1}^N$ denote the curvature
perturbation from projection of isocurvature perturbation and from
$\delta N$ respectively.

There are two kinds of non-Gaussianities in multi-stream inflation.
Firstly, the one-point probability distribution function for
$\zeta_{k_1}$ is non-Gaussian. This is because the extra density
fluctuation $\delta\zeta_{k_1}^N$, coming from the e-folding number
difference between trajectories $A$ and $B$, peaks at two particular
values:
\begin{align}\label{PN}
  P(\delta\zeta_{k_1}^N)=\frac{1}{2} \left[
\delta\left(\delta\zeta_{k_1}^N+\frac{1}{2}\delta
  N\right)+\delta\left(\delta\zeta_{k_1}^N-\frac{1}{2}\delta
  N\right)\right]~.
\end{align}
Here we use $P(x)$ to denote the probability distribution function.
This should not be confused with the power spectrum $P_\zeta$.
The contribution from the entropy perturbation $\delta\zeta_{k_1}^S$ is
still Gaussian.

Note that here we assumed for simplicity that the probability to pick
up trajectory $A$ and $B$ are equal. This seems natural in our
discussion because we are talking about a spontaneous breaking of an
approximate symmetry. One can generalize the discussion
straightforwardly to the case with unequal probability between
trajectories. With unequal probability, we can also provide a
theory for the CMB cold spot in the multi-stream inflation framework
\cite{Inprep}.

Note that the one point distribution function is not very easy to
probe, especially for one particular value $k_1$. So we now proceed
to discuss the other kind of non-Gaussianity produced by
multi-stream inflation. The idea is that the amount of perturbation
at comoving wave number $k_1$ and the amount of perturbation at
comoving wave number $k_1<k<k_2$ are correlated. The correlation
comes from the fact that the following three effects are $100\%$
correlated with each other: 1. An isocurvature perturbation deciding
whether to go along $A$ or $B$. 2. The difference between
$P_\zeta^{A}$ and $P_{\zeta}^B$. 3. The difference of the e-folding
number $\delta N$ between $A$ and $B$.

The three point correlation function can be written as
\begin{align}
  \langle \zeta_{k_1}\zeta_k\zeta_k \rangle = \langle \zeta_0\zeta_k\zeta_k \rangle+\langle \delta\zeta_{k_1}^S\zeta_k\zeta_k \rangle+\langle \delta\zeta_{k_1}^N\zeta_k\zeta_k \rangle~.
\end{align}
As we have neglected non-Gaussianities from sources other than the
multi-stream effect, we have
\begin{align}
 \langle \zeta_0\zeta_k\zeta_k \rangle=0~.
\end{align}
The two terms from entropy modes and $\delta N$ can be written as an
integration over the product of the three points and the probability
distribution function,
\begin{align}\label{corint}
  \langle \delta\zeta_{k_1}^{S,N}\zeta_k\zeta_k \rangle=
\int_{-\infty}^{\infty} d\delta\zeta_{k_1}^{S,N}
  d\zeta_k P(\delta\zeta_{k_1}^{S,N},\zeta_k) \delta\zeta_{k_1}^{S,N}\zeta_k\zeta_k~.
\end{align}
with the combined probability distribution function of
$\delta\zeta_{k_1}^{S,N}$ and $\zeta_k$,
\begin{equation}
  P(\delta\zeta_{k_1}^{S,N},\zeta_k)=P(\delta\zeta_{k_1}^{S,N})
  \left[
\frac{e^{-\frac{\zeta_k^2}{2\sigma^2_A}}}{\sqrt{2\pi}\sigma_A}\theta(\delta\zeta_{k_1}^{S})
+\frac{e^{-\frac{\zeta_k^2}{2\sigma_B^2}}}{\sqrt{2\pi}\sigma_B}\theta(-\delta\zeta_{k_1}^{S})
  \right]~,
\end{equation}
where $P(\delta\zeta_{k_1}^{S})$ takes a form of Gaussian
distribution
$\frac{1}{\sqrt{2\pi}\sigma_1^S}\exp\left\{-\frac{\left(\delta\zeta_{k_1}^S\right)^2}{2\left(\sigma_1^S\right)^2}\right\}$,
and $P(\delta\zeta_{k_1}^{N})$ is given in Eq. \eqref{PN} with
$\left(\sigma_1^S\right)^2=\langle \left(\zeta_{k_1}^S\right)^2
\rangle$, $\sigma_A^2=\langle \zeta_{k}^2 \rangle_A$ and
$\sigma_B^2=\langle \zeta_{k}^2 \rangle_B$. $\theta(x)$ is the
Heaviside step function. To avoid $\delta$ functions, we use the
normalization that the fluctuation is averaged over one Hubble
volume. (A familiar example is that in this normalization, the
fluctuation of the inflaton takes the form $\delta \varphi\sim
\frac{H}{2\pi})$.

One can perform the integration in Eq. \eqref{corint} and get
\begin{equation}\label{ngesti}
  \langle \zeta_{k_1}\zeta_k\zeta_k \rangle \simeq x P_{\zeta}^{1/2}
  \left(P_{\zeta}^A - P_{\zeta}^B\right)~,
\end{equation}
where $x\equiv \delta\zeta_{k_1}/\zeta_{k_1}$ denotes the fraction
of extra fluctuation from the multi-stream effect.

Now we compare the above result with the local form non-Gaussianity
to estimate $f_{NL}$ in the multi-stream model. The local form
$f_{NL}$ is defined as
\begin{equation}
\zeta=\zeta_g+\frac{3}{5}f_{NL}(\zeta_g^2-\langle
\zeta_g^2\rangle)~,
\end{equation}
or we can rewrite the above equation as
\begin{equation} \label{fnldef}
  f_{NL} \simeq \frac{\langle\zeta^3\rangle}{
  \langle\zeta^2\rangle\langle\zeta^2\rangle}~,
\end{equation}

Comparing Eqs. \eqref{ngesti} and \eqref{fnldef}, we have
\begin{equation}\label{fnlcal}
  f_{NL}\simeq x P_\zeta^{-1/2}\left( \frac{P_\zeta^A-P_\zeta^B}{P_\zeta}\right)~.
\end{equation}
Note that $P_{\zeta}^{-1/2}\simeq 10^{5}$ is a large number. We can
have $f_{NL}\simeq 100$ when $(P_\zeta^A-P_\zeta^B)/P_\zeta^A\simeq
3\times 10^{-3}$ and $x\simeq 0.3$. As we shall show, this needs a
tuning of the shape of the potential of order $1\%$. For
$f_{NL}\simeq 10$, the tuning is of order $10\%$. Note that this
non-Gaussianity appears only when one wave number is near $k_1$
(``near'' denotes about one e-fold during inflation), $f_{NL}$ is
strongly scale dependent, and should have large running.

We have not calculated the shape of non-Gaussianity for multi-stream
inflation in the present paper. However, one note that
non-Gaussianity is produced when $k\gg k_1$, known as the squeezed
limit in the literature. So we expect that the shape of
non-Gaussianity should be similar to the local shape, with large
running. So the local form estimator $f_{NL}$ should apply for our
model.

 Finally, let us
discuss the sign of $f_{NL}$. In multi-stream inflation, the sign of
$f_{NL}$ is determined so that a positive $\delta\zeta_{k_1}$
results in a larger perturbation at scale $k_1<k<k_2$. For example,
when a fluctuation $\delta\zeta_{k_1}>0$ decides that the field roll
to trajectory $A$, with $\zeta_A>\zeta_B$, then $f_{NL}>0$.

\begin{figure}
  \center
  \includegraphics[width=0.7\textwidth]{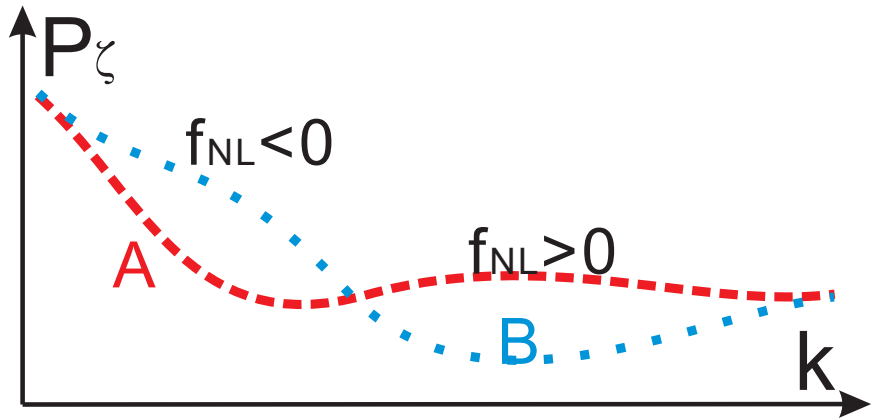}
    \caption{\label{ngrun} We illustrate the scalar power spectrum as a function of scale $k$.
The red (dashed) line denotes the trajectory $A$, and the blue
(dotted) line denotes the trajectory $B$. If a perturbation
$\delta\zeta_{k_1}>0$ leads to path $A$, then we have $f_{NL}<0$ on
larger length scales and $f_{NL}>0$ on smaller length scales.}
\end{figure}

It is also interesting to note that in the multi-stream model,
$f_{NL}$ can usually cross zero. It is because $P_\zeta^A$ may, say,
larger than $P_{\zeta}^B$ on some scales, than become smaller than
$P_{\zeta}^B$ on some other scales. As an example, when the
deformation of the potential is ``symmetric'',
\begin{equation}
  \int_{\varphi_1}^{\varphi_2}\delta V d\varphi=0~,
\end{equation}
and the slow roll parameters satisfy ${\cal O}(\epsilon)> {\cal
O}(\eta^2,\xi^2)$. In the lowest order of slow roll approximation,
one have
\begin{equation}
  \int_{\varphi_1}^{\varphi_2}\ d\varphi \left(P_\zeta^{A}
  -P_\zeta^{B}\right)=0~.
\end{equation}
So in this example, $f_{NL}$ must change sign at a certain scale.
This special kind of running of the non-Gaussianity is illustrated
in Fig.\ref{ngrun}

\section{Summary of Signatures for Multi-Stream Inflation}

In this section, we will give a summary of signatures for
multi-stream inflation.
As we have mentioned in Section 2, $\delta\zeta_{k_1}$ appears as an
additional power at scale $k_1$ in the CMB. This additional power
can not be arbitrarily large, because otherwise it should have been
observed in the experiments. To be in accordance with CMB experiments,
we require this anomaly power to satisfy $\delta\zeta_{k_1} < x_m
P_\zeta ^{1/2}$, where $x_m$ is determined by the error bar of CMB
experiments, which varies when we choose $k_1$ to be different
scales. Usually, one needs some fine tunning to satisfy the above
constraint.  In this
section, we also discuss this tuning issue in more detail.

First of all, the amount of tuning depends on the detail of the
potential, thus depends on the detail of string landscape. For
example, if the trajectories $A$ and $B$ are completely symmetric,
then there is no tuning at all. However, in this case, there are no
new signatures for multi-stream inflation either.

To analyze the relation between tuning and signatures for
multi-stream inflation further, let us calculate the relation
between $\delta N$ and $\delta V$. We use the slow roll parameters
\begin{equation}
  \epsilon\equiv \frac{M_p^2}{2}\left(
  \frac{V'}{V}\right)^2~, \qquad \eta\equiv
  M_p^2\frac{V{''}}{V}\qquad \xi^2\equiv M_P^4
  \frac{V' V{'''}}{V^2}~.
\end{equation}
We first investigate the parameter regime where $\epsilon$ is not
too small. To be more explicit, ${\cal O}(\epsilon)> {\cal
O}(\eta^2,\xi^2)$, and with ``symmetric'' deformation of potential:
\begin{equation}
  \int_{\varphi_1}^{\varphi_2}\delta V d\varphi=0~.
\end{equation}
With these assumptions, $\delta N$ can be written as
\begin{equation}\label{srest}
  \delta N=\int_{\varphi_1}^{\varphi_2} d\varphi \left(\frac{V+\delta V}
  {M_p^2(V'+\delta V')}- \frac{V}
  {M_p^2 V'}\right)\simeq \int_{\varphi_1}^{\varphi_2} d\varphi
  \frac{V}{M_p^2 V'}\left(\frac{\delta V}{V}-\frac{\delta V'}
  {V'}\right)~.
\end{equation}
The integration in Eq. \eqref{srest} can be expanded in powers of
the slow roll parameters. So the natural value of $\delta N$ is
\begin{equation}\label{deltaNest}
  \delta N_{\rm natural} \simeq (N_1-N_2)^2\left(
  -4\eta+\frac{3\eta^2}{\epsilon}-\frac{\xi^2}{\epsilon}\right)
   \frac{\delta V}{V}~.
\end{equation}
We find that $\delta N$ is suppressed by slow roll parameters. Note
that we have assumed $\int \delta V (\varphi-\varphi_1) d\varphi
\simeq (\varphi_2-\varphi_1)^2\delta V$. If for some reasons we have
$\int \delta V (\varphi-\varphi_1) d\varphi \ll
(\varphi_2-\varphi_1)^2\delta V$, then $\delta N$ can be much
smaller than given in Eq. \eqref{deltaNest}.


To summarize the experimental signatures for multi-stream inflation,
we propose two inequalities to divide the parameter space of multi-stream inflation
into four regions.

As discussed at the beginning of this section, current CMB experiments do not
allow large features. A condition $\delta\zeta_{k_1} < x_m
P_\zeta ^{1/2}$ is needed for multi-stream inflation to be consistent
with CMB observations. If the above inequality is nearly saturated,
then CMB experiments in the near future can detect the CMB features at scale $k_1$.

Another inequality comes from the comparison of $\delta V/V$ and $P_\zeta^{1/2}/x$, where
$x\equiv{\delta \zeta_{k_1}/P_\zeta^{1/2}}$. As shown in Eqs. \eqref{PV} and \eqref{fnlcal},
this comparison determines whether $f_{NL}$ is larger or smaller than ${\cal O}(1)$,
thus detectable or not in the near future.

There is no estimator as sharp as $f_{NL}$ for the power asymmetry. So before explicitly fitting
the multi-stream inflation model with CMB data, it is less clear how large power asymmetry
can be detected in the next generation of experiments. However, one should note that the detection
of $f_{NL}$ in Eq. \eqref{fnlcal} indicates a difference between $P_\zeta^A$ and
$P_\zeta^B$. So here, we take the existence of large non-Gaussianity to be the condition for
observable amount of power asymmetry. One should note that the detection of power asymmetry is
in principle easier than the detection of $f_{NL}$, because the detection of power asymmetry
does not need information for $\zeta_{k_1}$. We are glad to see whether improved methods
can be proposed for detecting the power asymmetry.

Using the above two inequalities, the four parameter regions for multi-stream inflation are:

\begin{itemize}
    \item When $\delta\zeta_{k_1} \ll x_m P^{1/2}_\zeta$ and $\delta V/V <
    P^{1/2}_\zeta/x$,
    there is no observable signature for multi-stream inflation. There
    may be an isocurvature perturbation, which is different from single
    field inflation. However, this isocurvature perturbation can
    also arise in simpler two field models.

    \item When $\delta\zeta_{k_1} \lesssim x_m P^{1/2}_\zeta$ and $\delta V/V <
    P^{1/2}_\zeta/x$,
    there is one signature for multi-stream inflation: a feature in the
    scalar power spectrum at scale $k_1$. Such features in the scalar power
    spectrum may be responsible for the features in the WMAP data.

    \item When $\delta\zeta_{k_1} \lesssim x_m P^{1/2}_\zeta$ and $\delta V/V >
    P^{1/2}_\zeta/x$, there are three signatures for multi-stream
    inflation: The first signature is as that in the above case:
    the feature in the spectrum at scale $k_1$. The second signature
    is that there will be an asymmetry in the powers between $k_1<k<k_2$
    in one patch of the sky of scale $k_1$ and another. This could
    be
    responsible for the hemispherical power asymmetry. The third prediction is
    non-Gaussianity. The non-Gaussianity is of order
    \begin{equation}
      f_{NL}\simeq x P_\zeta^{-1/2}\frac{\delta V}{V}>1~.
    \end{equation}
    As discussed in the former section, $f_{NL}$
    in our multi-stream model is largest in the squeezed limit
     because $k_1 \ll k$, and
     $f_{NL}$ usually crosses zero.

    Now we estimate whether this case is natural or not.
    The natural value of $\delta N$ can be estimated using Eq.
    \eqref{deltaNest},
    \begin{equation}
      \delta N_{\rm natural}\simeq (N_1-N_2)^2 {\cal O}(\epsilon)
      P_\zeta^{1/2} f_{NL}/x~,
    \end{equation}
    In order that perturbation on the scale $k_1$ is not
      ruled out by experiments, we
    need $\delta\zeta_{k_1}<x_m P_\zeta^{1/2}$. The amount of tuning for
    this parameter region is measured by
    \begin{equation}
    \frac{\delta \zeta_{k_1}}{\delta N_{\rm natural}}\simeq
      \frac{\delta N}{\delta N_{\rm natural}}\simeq \frac{x_m^2}
      {(N_1-N_2)^2 {\cal O}(\epsilon) f_{NL}}~.
    \end{equation}
    For example, if we have multi-stream inflation for 3 e-folds, $\epsilon =
    {\cal O}(0.01)$, and $x_m^2=0.1$, we have $\delta \zeta_{k_1}/\delta N_{\rm natural}
    \simeq 1/f_{NL}$. To be more explicit, there is one potential in
    10 potentials
    that can produce $f_{NL}=10$, and one in 100 that can produce
    $f_{NL}=100$. Assume that future experiments show $f_{NL}=10$,
    then 3 features in the scalar power spectrum can naturally
    achieve that.

    \item When $\delta\zeta_{k_1} \ll x_m P^{1/2}_\zeta$ and $\delta V/V
    >P^{1/2}_\zeta/x$, there is large non-Gaussianity, asymmetry on
    powers in different patches of the sky, and no feature in the
    power spectrum. However, more fine tuning is needed for
    this possibility than the former one.
\end{itemize}

There is potentially one more signature of multi-stream inflation:
the correlation between the curvature perturbation and the
isocurvature perturbation. However, as we need to go to detailed
models to calculate this correlation, we shall not consider it in
the present work. We hope to address this issue in the future.

If $\epsilon$ is even smaller than $\eta^2$ and $\chi^2$, then we
are unable to perform the expansion \eqref{deltaNest} in the slow
roll parameters. In this case, $\delta N/N \sim \delta V/V$, and we
need more fine tuning to get the desired non-Gaussianity and
anisotropy for perturbation powers for multi-stream inflation.

\section{Conclusion}
To conclude, we have described the multi-stream inflation model, and
calculated the density perturbation and non-Gaussianity of the
model.

We find that the model can be parameterized by the $\delta V/V$ and
$\delta N$. We show that signatures such as non-Gaussianity,
features in the CMB and the hemispherical power asymmetry can be
produced in
corresponding parameter regions.

We also note that isocurvature perturbation at $\varphi_1$
and $\varphi_2$ may produce interesting physics. We hope we can
address these issues in some future work. We also hope to build more
explicit models of multi-stream inflation, and investigate the
signatures of multi-stream inflation in more detail in the future.
Finally, the perturbations are studied semi-quantitatively in this
paper. It is important to calculate the perturbations more
precisely and determine the ${\cal O}(1)$ coefficients. To do so, one needs
to investigate the cosmic perturbation theory for a quantum state with two decohered branches. We also leave
this calculation to future work.

\section*{Acknowledgments}
We thank Bruce Bassett, Robert Brandenberger, Yifu Cai, Xingang Chen, Qing-Guo Huang and Xian Gao for discussion. This work was supported by NSFC
Grant No. 10525060, and a 973 project grant No. 2007CB815401.

\end{document}